\documentclass[11pt, article]{amsart}

\usepackage{amsmath}
\usepackage{latexsym}
\usepackage[all]{xy}
\usepackage{color}
\newtheorem{teorema}{Theorem}[section]
\newtheorem{definicion}[teorema]{Definition}
\include{amslatex}
\newtheorem{proposicion}[teorema]{Proposition}

\usepackage[margin=1.5in]{geometry}

 \numberwithin{equation}{section}

\begin{document}
\begin{title}[Effect of the maximal proper acceleration in the inertia]
{On the effect of the maximal proper acceleration in the inertia}
\end{title}
\maketitle
\begin{center}
\author{Ricardo Gallego Torrom\'e}\footnote{Email: rigato39@gmail.com}
\end{center}
\begin{center}
\address{Department of Mathematics\\
Faculty of Mathematics, Natural Sciences and Information Technologies\\
University of Primorska, Koper, Slovenia}
\end{center}

\begin{abstract}
The effect of a hypothetical maximal proper acceleration on the mass of a charged particles is investigated in the context of particle accelerators. In particular, it is shown that maximal proper acceleration implies an increase in the kinetic energy of the particle being accelerated with respect to the relativistic energy. Such an increase in kinetic energy leads to a reduction of the luminosity of the bunches with respect to the expected luminosity in the relativistic models of the bunches. This relative loss in luminosity is of the order $10^{-3}$ to $10^{-5}$ for the LHC bunches and can be of order up to $10^{-3}$ for certain laser plasma accelerator facilities. Although the effect is small, it increases with the square of the bunch population.
\end{abstract}
\section{Introduction}
The problem of finding a classical, consistent dynamical model for the point charged particle has attracted the attention of researchers for a long time, not only because the need of having a consistent theory of classical electromagnetism where radiation reaction effects are consistently described \cite{Dirac, Dirac1951}, but also as an attempt to have a framework for quantization that could avoid the problems of infinities in the quantization of a classical field theory of electrodynamics \cite{BornInfeld, BornInfeld2,BornInfeld3}.
Most of the efforts to construct an improved theory of classical electrodynamics have been developed in the framework where the electromagnetic field is represented as a smooth section of a tensor bundle over a spacetime manifold endowed with a Lorentzian structure. Lorentzian geometry has been outstandingly successful in the description of the physical spacetime arena in relativistic theories and constitutes one of the pillars of modern theoretical physics. However, the logical foundations of Lorentzian geometry as a fundamental block of physics rely on several assumptions that, from a formal point of view, remain still not completely justified. This is especially the case of the {\it clock hypothesis} in relativity theory, the requirement that the proper time functional associated to the spacetime metric depends formally only on the world line curve and on the tangent vector along the curve where the proper time functional is being evaluated \cite{Einstein1922, Pauli 1958}. Indeed, when only consistence with the clock hypothesis is required, the natural geometric arena is not the category of Lorentzian geometry, but the category of pseudo-Finsler geometry, where the metric depends on the position and velocity vector of worlds lines \cite{Beem1,GallegoPiccioneVitorio:2012}. Furthermore, there are also theoretical reasons to believe that the clock hypothesis fails in situations when radiation reaction effects are of relevance, in particular in situations where the particle experiences large accelerations \cite{Mashhoon1990}.

The above arguments provide theoretical motivation to consider spacetime structures that could depend upon higher order derivatives, and in particular, on acceleration fields along world lines. One natural way to describe such a dependence is by means of a small, linear perturbation acceleration depending on the original Lorentzian structure. The {\it smallness} of the perturbation implies the existence of a well-defined connection theory, bearing a close similarity to the theory of Randers  spaces in Finsler geometry. Such structures contain restrictions on the acceleration. In particular, there is an upper, uniform bound for the proper accelerations \cite{Ricardo2015}.

In connection with the above generalization of the metric structure of the spacetime, a research program aimed to developed a mathematical theory of fields that depend upon higher order jets was developed motivated by  the following insight. In theory, the measurement of an electromagnetic field is done by observing the results of the scattering of test charged particles with the given field. It is usually assumed in classical field theory that the effect of the measurement does not disturb the value of the field. But this is not necessarily true if the accelerations of the test particles involved are large. In such situations, the effect of measuring the field could induce a significant change in the field being measured, due to radiation by the charged test particle. This intrinsic feed-back problem of measuring the electromagnetic field can be overcome by assuming that fields are objects that depend on how they are measured, that is, on the particularities of the kinematics of the test particle's world line. A natural way to implement mathematically this idea is by means of higher order jet field theory. Thus in the case of electrodynamics, electromagnetic fields and spacetime metric geometry will be objects depending on higher order jets. On the other hand, if fields depend on higher order jets and the spacetime geometry is controlled by the matter content, then also the spacetime geometry must depend on higher order jets. But as we have mentioned above, a natural way to describe this type of geometry is by means of a small, linear perturbation of a Lorentzian metric, where the perturbation depends upon the acceleration. Thus the assumption of an acceleration dependent metric acquires full sense in the setting of higher order jet fields models. Motivated by these insights, a geometric setting of fields depending on higher order jets was developed in \cite{Ricardo2012} and led to a classical second order ordinary differential equation model for the point charged particle that does not have pre-accelerated and run away solutions \cite{Ricardo2015, Ricardo 2017}.

The present article explores the effect of maximal proper acceleration on the mass and kinetic energy of point charged particles in the context of a theory of higher order jet electrodynamics and spacetimes of maximal proper acceleration. The possibility of a relation between acceleration and  mass is not novel. J. Larmor \cite{Larmor} and later W. Bonnor \cite{Bonnor} suggested the possibility that the inertial mass could depend upon the proper time of a relativistic metric as the resource for the origin of the radiation of energy of the particle. In particular, a second order differential equation for the point electron was proposed by Bonnor in a theory where the inertial mass depends on the acceleration in the framework of Minkowski spacetime \cite{Bonnor}. Even though this sector of Bonnor's model is formally identical to our equation of motion for the point charged particle derived, both models differ in  radical ways, especially in the framework where the two theories are developed and in the nature of the inertial mass coefficient that appears in the equation of motion. In our theory, inertial mass and rest mass are different concepts and while in Bonnor's theory the inertial mass depends on proper time according to a differential equation law, in our model the inertial mass is constant, but the rest mass depends on the acceleration and therefore on the time by means of an algebraic expression. Furthermore, in our theory, the rest mass or the inertial mass are not the resource for the energy radiated by the particle. Instead, the unifying element field-test particle is conveyed by the notions of a higher order jet electromagnetic field.

The structure of this paper is the following. First, in {\it section} \ref{Maximal acceleration spacetimes and its kinematics} we have recalled certain fundamental notions from the theory of metrics of maximal acceleration that are need to formulate the concepts involved in our discussion. Then the notion of $4$-momentum and the relation of mass with acceleration in a spacetime with maximal acceleration are discussed, leading to the introduction of the notion of rest mass in our theory. After this, the dynamical model of point charged particle that we use is introduced, with special attention to the case of bunches of particles moving coherently in accelerator systems. In {\it section} \ref{Model of Bunch} the model of the bunch that we use has been explained. The bunch of particles is treated as a point charged particle whose mass is the total mass of the bunch and whose charge is the total charge of the bunch. Although this model is an over-simplified description of the bunch, it is enough to show theoretical evidence of observable effects due to the increase of the kinetic energy of the particle with respect to the relativistic kinetic energy due to the effect of maximal proper acceleration, there must be a decrease in the particle population of the bunches. In {\it section} \ref{Effects} these effects on the kinematics and population have been discussed and illustrated with several examples of particle accelerator systems. In particular,
we have considered the case of the LHC bunches and several laser-plasma and plasma wakefield acceleration cases. For the LHC, the increase in the relativistic energy is in the range $10^{-3}$ to $10^{-5}$ (depending on the energy of the particles). For the LHC, the effect is mainly due to the acceleration in the bending magnets, because the bunch passes many times through dipole magnets compared with the time it passes by radio frequency cavities and because the proper acceleration in the dipole magnets is relatively higher than in the radio frequency systems.
For laser-plasma accelerator systems, assuming that the self-interacting effects do not affect the stability of the bunch, the effect can be of order $10^{-3}$ to $10^{-4}$ relative to the standard relativistic energy of the bunch. This range depends on the accelerator facility in question. Also note that the proper acceleration reached is several orders of magnitude higher than the reached at LHC, but the bunch is accelerated only one time, which can render the effect more visible for observational purposes.

{\it Section} \ref{Conclusion and Discussion} contains a discussion of the effect of space charge and beam current in our predictions. We conclude that, as long as very general bunch stability conditions hold good and we are in the regime of validity of our analysis, the prediction of the order of magnitude of the effects discussed will not be altered.
Finally, we have also discussed the notion of higher order fields and the generalization of exterior derivatives and start product operations to higher order jet differential algebra   in {\it appendix} \ref{Appendix A}, where also the generalized linear electrodynamics that we use has been briefly introduced. In {\it appendix} \ref{Appendix B} the notion of spacetime of maximal proper acceleration in the context of higher order jet geometry has been introduced. Finally, the derivation of the equation of motion of a point particle in the framework of classical electrodynamics of higher order jet fields in a spacetime of maximal acceleration has been summarized in {\it appendix} \ref{Appendix C}. Further details regarding these concepts and on related developments can be found in references \cite{Ricardo2015,Ricardo2012,Ricardo 2017, Ricardo2007, Ricardo 2020}.
\section{Spacetimes of maximal proper acceleration and its kinematics}\label{Maximal acceleration spacetimes and its kinematics}
\subsection{Spacetimes of maximal proper acceleration as high order jet geometries}
We start by considering the construction of the metric of maximal acceleration (see {\it appendix} \ref{Appendix B} for a discussion of the notion of spacetime of maximal acceleration from the point of view of higher order jet geometry). The assumption that we make is that the spacetime structure is of the form $g=\,\eta+\,\delta g$.
The leading order term of a spacetime of maximal proper acceleration $(M_4,g)$ is a Lorentzian structure $(M_4,\eta)$, where the metric $\eta$  has signature $(-1,1,1,1)$. The Levi-Civita connection of $\eta$ is denoted by $\nabla$. The metric $\eta$ is interpreted as the limit metric of $g$ when the test particles do not experience acceleration. In the case of electrodynamics, this can be the case either because the test particles are not charged or because there is no external electromagnetic field in the spacetime region under consideration.
 Let us consider a smooth curve $\vartheta:I\to M_4$ whose tangent vector field is time like with respect to the limit Lorentzian metric $\eta$, $\eta({\vartheta}'(t),{\vartheta}'(t))<0$,
where ${\vartheta}'(t)$ is the tangent vector of the curve $\vartheta:I\to M_4$ parameterized by an arbitrary parameter $t\in\,I\subset\,\mathbb{R}$. The proper time of the curve  $\vartheta:I\to M_4$ with respect to $\eta$ is
\begin{align}
\tau [\vartheta] :=\,\int_{I} \,dt\,\left( -\eta({{\vartheta}'},{{\vartheta}'})\right)^{1/2} .
\label{propertime eta}
\end{align}

With this geometric input, we can define a non-degenerate, symmetric form $g$ such that when probed by a test particle with world line $\vartheta:I\to M_4$ it is given by the expression
\begin{align}
g(\,j^2{\vartheta}(t)) (X,Y):=\Big(1+ \frac{  \eta(\nabla_{{\vartheta}'}{\vartheta}'(\tau)
\nabla_{{\vartheta}'}{\vartheta}'(\tau))}{A^2 _{\textrm{max}}\,\eta({\vartheta}',{\vartheta}')}\Big)\, \eta (X,Y),\quad\quad X,Y\in \,\Gamma TM,
\label{maximalaccelerationmetric}
\end{align}
where $j^2({\vartheta}^\mu(t))=\,(\vartheta^\mu(t),\frac{d}{d t}\vartheta^\mu(t),\frac{d^2}{d t^2}\vartheta^\mu(t))$.

We assume that:
 \begin{itemize}
 \item The time like character of curves is the same for $g$ and for $\eta$:
 \begin{align*}
 g({\vartheta}',{\vartheta}')<0\,\Leftrightarrow\,\eta({\vartheta}',{\vartheta}')<0,
\end{align*}
for every time like curve $\vartheta:I\to M_4$.
\item The covariant condition
\begin{align}
\eta(\nabla_{{\vartheta}'}\,{\vartheta}',\,\nabla_{{\vartheta}'}\,{\vartheta}')\,\geq 0.
\label{spacelikeaccelerations}
\end{align}
holds good.
 \end{itemize}
  These two assumptions ensure the existence of a causal structure for $(M_4,g)$ and a well-defined connection theory for $g$ \cite{Ricardo 2020}. Moreover,
assuming the above assumptions, the bound
 \begin{align}
  0 \leq g(\nabla_{{\vartheta}'}{\vartheta}',\nabla_{{\vartheta}'}{\vartheta}')\leq\,\eta(\nabla_{{\vartheta}'}{\vartheta}',\nabla_{{\vartheta}'}{\vartheta}')
  <\,A^2_{\textrm{max}}\,\left(-\eta({\vartheta}',{\vartheta}')\right)^{1/2}
  \label{boundedconditionforacceleration}
 \end{align}
 holds good.

 Let us consider the case when the world line $\vartheta$ is parameterized by the proper time of $\eta$, which is given by the expression \eqref{propertime eta}. Derivatives with respect to $\tau$ are denoted by dot notation: $\dot{\vartheta}(\tau):=\frac{d}{d\tau}\vartheta(\tau)$, etc... By a maximal proper acceleration $A_{\textrm{max}}$ we mean that the bound \eqref{boundedconditionforacceleration} for the value of the squared proper acceleration $a^2:=\,\eta(\nabla_{\dot{\vartheta}}\dot{\vartheta},\nabla_{\dot{\vartheta}}\dot{\vartheta})$ holds good,
 \begin{align}
   0 \leq \eta(\nabla_{\dot{\vartheta}}\dot{\vartheta},\nabla_{\dot{\vartheta}}\dot{\vartheta})<\,A^2_{\textrm{max}}.
  \label{boundedconditionforaccelerationtau}
 \end{align}

 Finally, let us remark that $\tau [\vartheta] $ is equal or larger than the proper time of the metric of maximal acceleration, which is defined by the expression
\begin{align}
s [\vartheta] :=\,\int_{\tilde{I}} \,d\tau\,\left( -g(\dot{\vartheta},\dot{\vartheta})\right)^{1/2} = \,\int_{\tilde{I}} \,d\tau\,\left(1-\frac{\eta(\nabla_{\dot{\vartheta}}\dot{\vartheta},\nabla_{\dot{\vartheta}}\dot{\vartheta})}{A^2_{\textrm{max}}}\right)^{1/2},
\label{propertime g}
\end{align}
where $\tau\in\,\tilde{I}$.
 Note that the functional $s [\vartheta]$ is not re-parametrization invariant and that in the expression \eqref{propertime g}, the parametrization of the curve is the proper time of the metric $\eta$.

\subsection{Notions of $4$-velocity in a geometry of maximal acceleration}
Given the precedent geometric constructions, the $4$-velocity vector field along the curve $\vartheta: I\to M_4$ is given at the instant $t$ by the expression
\begin{align}
v^\mu(t)=\,\frac{1}{\sqrt{1-\frac{a^2(t)}{A^2_{\textrm{\textrm{max}}}}}}\,\tilde{v}^\mu(t),\quad \mu=1,...,4,\quad t\in \,I,
\label{fourvelocity}
\end{align}
where $\tilde{v}(t)$ is the standard $4$-velocity vector and where the value of the acceleration squared $a^2 (t) :=\eta(\nabla_{\dot{\vartheta}}\,\dot{\vartheta},\,\nabla_{\dot{\vartheta}}\,\dot{\vartheta})$ is evaluated using the parameter $\tau$ instead of $t$.

When the  metric $\eta$ is the Minkowski metric, in a inertial coordinate system it has diagonal form  $\eta=diag\,(-1,1,1,1)$ and the relativistic $4$-velocity $\tilde{v}^\mu(t)$ is related with the Newtonian velocity $\vec{\bf {v}}$ by the expression
\begin{align*}
 \tilde{v}^0(t)=\,\frac{1}{\sqrt{1-\frac{\vec{\bf {v}}^2(t)}{c^2}}}\,c,\quad
\vec{\tilde{v}}(t)=\,\frac{1}{\sqrt{1-\frac{\vec{\bf {v}}^2(t)}{c^2}}}\,\vec{\bf {v}}(t).
\end{align*}
Then  from the expression \eqref{fourvelocity}, one has the following relations for the components of the $4$-velocity $v^\mu(t)$,
\begin{align}
v^0(t)=\,\frac{1}{\sqrt{1-\frac{a^2(t)}{A^2_{\textrm{max}}}}}\frac{1}{\sqrt{1-\frac{\vec{\bf {v}}^2(t)}{c^2}}}\,c,\quad
\vec{v}(t)=\,\frac{1}{\sqrt{1-\frac{a^2(t)}{A^2_{\textrm{max}}}}}\frac{1}{\sqrt{1-\frac{\vec{\bf {v}}^2(t)}{c^2}}}\,\vec{\bf
{v}}(t) .
\label{componentsv0123}
\end{align}

\subsection{Redefinition of the energy and momentum observables} The coefficient $m_0$ that appears in the covariant form of Newton's second law of dynamics, $F^{\mu}= \,m_0 \,\ddot{x}^\mu,\, \mu=0,1,2,3$, is the {\it inertial mass} of the system.
\begin{definicion}
Let $(M,g)$ be a spacetime of maximal acceleration and $O\in\,\Gamma \,TM_4$ an observer. The $4$-momentum  of a point particle with inertial mass $m_0$ and world line
$\vartheta:I\to M$ observed by $O$ is defined by the components
\begin{align}
P(t)=\,m_0\,v(t),
\label{fourmomentum}
\end{align}
where $v$ is the $4$-velocity measured by $O$ and whose components are given by the expressions \eqref{componentsv0123}.
\end{definicion}

When $\eta$ is the Minkowski metric, inertial coordinate systems are defined. The components of the $4$-velocity in an inertial coordinate system are
\eqref{componentsv0123}. Then the components of the four-momentum are given by
 \begin{align}
c\,P^0(t)=E(t)=\,\frac{1}{\sqrt{1-\frac{a^2(t)}{A^2_{\textrm{max}}}}}\,\frac{1}{\sqrt{1-\frac{\vec{\bf {v}}^2(t)}{c^2}}}\,m_0c^2,
\label{modifiedE}
\end{align}
\begin{align}
\vec{\bf P}(t)=\,\frac{1}{\sqrt{1-\frac{a^2(t)}{A^2_{\textrm{max}}}}}\,\frac{1}{\sqrt{1-\frac{\vec{\bf {v}}^2(t)}{c^2}}}\,m_0\vec{\bf {v}}(t) .
\label{threemoment}
\end{align}

According to the above re-definition of energy of a point particle, in the instantaneous Lorentzian coordinate frame where the particle has associated the Lorentz factor $\gamma_{\vec{v}}$, the energy of the system is given by the expression \eqref{modifiedE}.
This re-definition of energy implies that the mass of an accelerated particle is
\begin{align*}
m (t) :=\,\frac{1}{\sqrt{1-\frac{\vec{\bf {v}}^2(t)}{c^2}}}\frac{m_0}{\sqrt{1-\frac{a^2(t)}{A^2_\textrm{max}}}}.
\end{align*}
In particular, in the instantaneously co-moving frame where $\vec{\bf v} =\vec{0}$ the {\it rest mass} is given by the expression
\begin{align}
m (t) =\,\frac{m_0}{\sqrt{1-\frac{a^2(t)}{A^2_\textrm{max}}}}.
\label{mass of an accelerated particle}
\end{align}
Thus the existence of a finite maximal proper acceleration implies that the rest mass \eqref{mass of an accelerated particle} increases with respect to the {\it inertial mass} $m_0$, which coincides with the rest mass of the particle only when the particle is not being accelerated.
\section{The model of the classical point charged particle in spacetimes of maximal acceleration}\label{Model of Bunch}
In the theory of higher order jet electrodynamics the spacetime structure that determines the proper time is a metric of maximal proper acceleration \eqref{maximalaccelerationmetric}.
Under this assumption and assuming compatibility with the covariant Larmor's radiation formula \cite{Jackson}, general arguments lead to a  model for the point charged particle described by a second order differential equation (see {\it appendix}  \ref{Appendix C} for a derivation of the equation of motion in the framework of generalized electrodynamics fields according to the theory developed in \cite{Ricardo 2017}). In a general covariant formalism, the equation of motion for a point charged particle of inertial mass $m_0$ and charge $q$ is expressed as
\begin{align}
m_0\,\nabla_{\dot{\vartheta}}\dot{\vartheta} =\,q\,\widetilde{\iota_{\dot{\vartheta}}F}-\,\frac{2}{3}\,{q^2}\,g(\nabla_{\dot{\vartheta}}\dot{\vartheta},\nabla_{\dot{\vartheta}}\dot{\vartheta})\,\dot{\vartheta},
\label{equationofmotion}
\end{align}
where, using the notation from \cite{BennTucker}, $F$ is the $2$-form Faraday form, $\iota_{\dot{\vartheta}}$ is the interior derivative with respect to $\dot{\vartheta}$ and $\widetilde{\iota_{\dot{\vartheta}}F}$ is the dual form determined by the metric $\eta$ of the  $1$-form ${\iota_{\dot{\vartheta}}F}$.

Some remarks related to equation \eqref{equationofmotion} are in order. First, equation \eqref{equationofmotion} is a second order differential equation with respect to the proper time parameter of $\eta$, in contrast with the Lorentz-Dirac equation, which is a third order equation \cite{Dirac}. Therefore, it is natural that equation \eqref{equationofmotion} avoids the pre-accelerated solutions that plague the Lorentz-Dirac equation, that has their origin in the third order derivatives that appear in it. Second, one can also show that the equation \eqref{equationofmotion} does not have run-away solutions \cite{Ricardo 2017}, a fact consistent with the assumption that the spacetime arena is a spacetime of maximal acceleration.

Equation \eqref{equationofmotion} implies a maximal proper acceleration for particles accelerated by means of electromagnetic fields. Such an acceleration scale can be obtained as follows. If in equation \eqref{equationofmotion} we contract the left side with the left side and the right side with the right side using the metric of maximal acceleration $g$, then it follows that
 \begin{align*}
 m^2_0\left(1-\frac{a^2}{A_\textrm{max}}\right)\,a^2\,= \,F^2_L +\,\left(\frac{2}{3}\,q^2\right)^2\,(a^2)^2\,g(\dot{\vartheta},\dot{\vartheta}).
 \end{align*}
 where
 \begin{align*}
 F^2_L =\, q^2\, g(\widetilde{\iota_{\dot{\vartheta}}F},\widetilde{\iota_{\dot{\vartheta}}F})\geq 0.
 \end{align*}
 Since $F^2_L$ is greater or equal to zero and $g(\dot{\vartheta},\dot{\vartheta})\geq \,\eta(\dot{\vartheta},\dot{\vartheta})= \,-1$, it follows that
  \begin{align*}
 m^2_0\,\left(1-\frac{a^2}{A^2_\textrm{max}}\right)a^2+\left(\frac{2}{3}\,q^2\right)^2\,(a^2)^2\geq 0.
 \end{align*}
Since $a^2 >0$, then we have that
 \begin{align*}
 1> \, a^2\left(\frac{1}{A^2_{\textrm{max}}}-\left(\frac{2 q^2}{3m_0}\right)^2 \right)
 \end{align*}
It follows the bound
\begin{align}
a^2\leq A^2_{\mathrm{max}}< \left(\frac{3}{2}\,\frac{m_0}{q^2}\right)^2 .
\label{bound in acceleration}
\end{align}
Therefore, the maximal acceleration of a point charged particle whose dynamics is determined by the equation \eqref{equationofmotion} and is given by the expression
\begin{align}
 A^2_{\mathrm{max}}=\,\left(\frac{3}{2}\,\frac{m_0}{q^2}\right)^2 ,
\label{valueofthemaximalacceleration}
\end{align}
since there is no other acceleration scale between $0$ and $\frac{3}{2}\,\frac{m_0}{q^2}$ in the model. It is because the specific dependence of the proper acceleration, namely, a dependence of the form $m_0/q^2_0$, that we will show it has phenomenological consequences for the luminosity of the bunches in particle accelerators.

In local Fermi coordinates of $\eta$ the equation of motion of a point charged particle \eqref{equationofmotion} takes the form
\begin{align}
m_0\,\ddot{x}^{\mu} =\,q\,F^{\mu}\,_{\nu}\,\dot{x}^{\nu}-
\,\frac{2}{3}\,{q^2}\,\ddot{x}^{\nu}\ddot{x}_{\nu}\,\dot{x}^{\mu},\quad F^{\mu}\,_{\nu}=\,g^{\mu\rho}F_{\rho\nu}, \quad \mu,\rho,\nu =0,1,2,3.
\label{equationofmotionnoncovariant expression}
\end{align}
Equation \eqref{equationofmotionnoncovariant expression} formally resembles the equation derived by W. Bonnor under the assumption that the energy radiated by a point charged particle according to Larmor's law has its origin in the inertial mass $m_0$ of the particle \cite{Bonnor}. However, Bonnor's theory and the theory that we are considering here differ considerably. Indeed, the spacetime framework in Bonnor's theory is Minkowski spacetime and the fields are Maxwell field, that is, classical fields solutions of Maxwell equations. In the relativistic framework, Bonnor's dynamical model consists of equation \eqref{equationofmotionnoncovariant expression} together with the condition for the inertial mass,
\begin{align}
\frac{d}{d\tau}{m_0} = \,-\frac{2}{3}\,q^2\,\dot{v}^\mu\dot{v}_\mu.
\label{Bonnor condition 2}
\end{align}
This change in the inertial mass compensates the change in four-momentum equal to $\frac{2}{3}\,q^2\,(\dot{v}^\rho\dot{v}_\rho){v}^\mu$ experienced by an accelerated charged particle due to the emission or absorbtion of radiation \cite{Jackson}. Note that the dependence on time of the inertial mass $m_0$ in Bonnor's theory given by the expression \eqref{Bonnor condition 2} is different than the dependence on time of the rest mass according to the definition \eqref{mass of an accelerated particle}. Thus one cannot identify inertial mass in Bonnor's theory with rest mass in our theory.
Also note that in Bonnor's theory there is no obvious uniform bound of the acceleration, since the inertial mass $m_0$, that could eventually appear in the expression for the maximal acceleration, depends on time.
\subsection{Model of the bunch of particles as a sole particle}
In order to investigate the main consequences of the expression \eqref{valueofthemaximalacceleration} in particle accelerators we use a simplified
model of the particle bunch, which has been previously discussed partially in \cite{Ricardo 2019}. We consider each individual bunch of particles as a single charged particle with  charge  $N q$ and inertial mass $N \, m_0$, where $q$ and  $m_0$  are the charge and the inertial mass of each individual particle composing the bunch and $N$ is the number of the particles of the bunch. This model contrasts with the rich dynamics of the accelerated bunch, which is usually described by means of fluid models, kinetic models, or statistical mechanical models \cite{Davidson Qin}, but it is useful to our aim of investigating the order of magnitude of the effects of maximal acceleration. The equation of motion \eqref{equationofmotion} can be treated perturbatively in terms of the quotient $\epsilon=a^2/A^2_{\textrm{max}}$, with the leading term being the Lorentz force term. Consequently, at zero order in $\epsilon$, the model for a charged particle is equivalent to the Lorentz force equation. Assuming this, the acceleration $|a|$ depends upon the quotient $m_0/q$, while the maximal proper acceleration  $A_{\textrm{max}}$ depends on the quotient $m_0/q ^2$. Thus under the influence of external fields of the accelerator and for acceleration squared $a^2 $ small enough compared with $A^2_{\textrm{max}}$, for the model of the bunch as a single particle with charge and mass $(N\,m_0, N\,q)$, the bunch will have the same acceleration as an individual point charged particle with charge and mass $(m_0, q)$.

Besides disregarding all the effects of beam dynamics related with the composition, size and structure of the bunch, especially disregarding space charge effects, the model of the bunch as a sole particle breaks down when
\begin{align*}
\big| F^2_L \big|\sim \,\big|\left(\frac{2}{3}\,q^2\right)^2\,(a^2)^2\,g(\dot{\vartheta},\dot{\vartheta})\big |,
\end{align*}
a condition that can be re-casted in a closer form:
\begin{align}
\left(\frac{q}{m}\right)^2\,\big| g(\widetilde{\iota_{\dot{\vartheta}}F},\widetilde{\iota_{\dot{\vartheta}}F})\big|\sim \,\frac{a^2}{A^2_{\textrm{max}}}\,a^2,
\label{consistence condition of the model}
\end{align}
where we have used the condition $\eta(\dot{\vartheta},\dot{\vartheta})=\,-1$. The left side of this condition does not depend upon the number population $N$. Thus, as long as we are far from reaching the condition \eqref{consistence condition of the model} and other dynamical effects are disregarded, the approximation of treating the bunch as a single charged particle with mass $N\, m_0$ and charge $N\, q$ remains valid.

\section{Phenomenological consequences of the redefinition of the energy in accelerator physics}\label{Effects}
The difference in the definition of energy $E$ between the theory with maximal proper acceleration
and the energy of the system according to the theory of relativity takes the form
\begin{align}
\Delta E =\,\gamma_{\vec{v}}\, \left(\frac{1}{\sqrt{1-\frac{a^2}{A^2_\textrm{max}}}}-1\right)\,m_0\,c^2 =\,\frac{1}{2}\,\gamma_{\vec{v}}\,\frac{a^2}{A^2_\textrm{max}}\,m_0\,c^2+\,\mathcal{O}(a^2/A^2_{\textrm{max}})^2,
\end{align}
where $\gamma_{\vec{v}}$ is the relativistic factor, $\gamma_{\vec{v}} =\,1/\sqrt{1-\frac{\vec{v}^2}{c^2}}$.
For $a^2/A^2_{\textrm{max}}\ll 1$, the first order approximation in $\epsilon$ is reasonable for our purposes. For a particle of charge and mass $(q,m_0)$, we have that the maximal acceleration is of the form \eqref{valueofthemaximalacceleration}. However, if the bunch being accelerated contains $N$ particles and if the bunch of particles is considered as a sole particle, then the maximal acceleration is decreased by a factor $N$. Hence the excess in energy with respect to the relativistic energy due to a maximal acceleration is given by the expression
\begin{align*}
\Delta E^{l=1} / E_{\textrm{rel}} = \, \frac{1}{2}\, N^2 \,\frac{a^2}{A^2_\textrm{max}(1)},
\end{align*}
where $A^2_\textrm{max}(1)$ is the maximal proper acceleration that applies to a sole particle.
If the systems passes by an identical acceleration device $M$ times, the additive effect leads to
\begin{align}
\Delta E^{l=M} / E_{\textrm{rel}} = \, \frac{1}{2}\, M\,N^2\,\frac{a^2}{A^2_\textrm{max}(1)}.
\label{effect of energy in the accelartion in an accelerator}
\end{align}
In terms of the value of the maximal acceleration we have
\begin{align}
\Delta E^{l=M} / E_{\textrm{rel}} = \, \frac{2}{9}\,\frac{q^4}{m^2_0} M\,N^2\,a^2 .
\label{effects depends upon the mass}
\end{align}
As an approximation, the value of $a^2$ can be calculated using the Lorentz force equation for the given external fields.

In order to illustrate the theory, let us consider the case of the effects of the existence of maximal acceleration at the LHC. In this case, the particles accelerated are protons. The maximal number of particles per bunch is of order $N=\,10^{11}$. The number of turns before the collisions is  $4 \times \,10^8$.  Since there are $8$ radio frequency (RF) cavities per direction, the number of independent accelerations is $M\sim \,3.2\times 10^{9}$. In the case of the RF-cavity acceleration, in the co-moving frame, the bunch experiences an electric field that is of order $|\vec{E}|=\,2 \times 10^6$ Volts/m.  Then the order of magnitude of the proper acceleration is obtained by applying the Lorentz force equation, $|\eta(a,a)|^{1/2}\sim  \frac{q}{m_0}\,|\vec{E}|\sim 2\times \,10^{15}\,m/s^2$. On the other hand, the maximal acceleration for an individual proton is $A_{\textrm{max}}(1)= \, \frac{3}{2}\frac{m_p}{e^2}\approx 10^{35} \,m/s^2$.  Thus we have that for the RF-cavity acceleration, the energy increase between the relativistic theory and the theory of electrodynamics in spacetimes of maximal acceleration that we are discussing is of order $\Delta E^{l=10^{9}} / E_{\textrm{rel}} \sim \,10^{-9}$.

The effect of the acceleration when the bunch passes through the electromagnetic field of the dipole magnets is estimated as follows. Each dipole magnet generates a maximal field of $8.3\, T$. By applying the Lorentz force as a first approximation, the acceleration of each proton in the bunch is maximally of order $a\sim \, {\gamma}\,\frac{q}{m_0}\,c |\vec{B}|\sim \,\gamma\,\cdot \,10^{14}\, m/s^2$. From injection, the relativistic gamma factor varies from $450$ to a maximum up to $7000$. Thus the proper acceleration in the dipole magnets is in the range $a=10^{16} \,m/s^2$ to $a=\,10^{17}\,m/s^2$. The number of dipole magnets is of order $1300$. Since the number of turns is $10^8$, this implies $M\sim 10^{11}$. For bunches with $N\sim\, 10^{11} $ protons, this leads to  a discrepancy in the energy with respect to the relativistic case in the range $\Delta E^{l=10^{11}} / E_{\textrm{rel}} \in\,[10^{-5},\,10^{-3}]$. This effect is larger than the one expected for the linear acceleration described above, but still small.

The increase in the energy of an accelerated bunch with respect to the relativistic energy implies a deficit in the bunch population $N$. This is because the energy injected in the dynamical system is fixed, but there is a larger rest mass due to the proper acceleration, according to the expression \eqref{mass of an accelerated particle}. The stability of the bunch of particles implies that this difference in energy is translated into a situation where fewer particles are actually accelerated to the required energy.
 If the energy introduced in the system is $E_T= E_{\textrm{rel}} N$, where $N$ is the expected number of particles in the bunch according to relativity and if equipartition of energy holds good, then equipartition in terms of the different population $N+\Delta N$ in our theory implies $
(E_{\textrm{rel}}+\Delta E) (N+\Delta N) = \,E_{\textrm{rel}} N $.
Thus we have that
\begin{align}
\frac{\Delta N}{N} =\,-\frac{\Delta E}{E_{\textrm{rel}}}=\,-\frac{2}{9}\,\frac{q^4}{m^2_0} M\,N^2\,a^2 .
\end{align}
For the example of the acceleration of magnets of proton bunches in the LHC as discussed above, we have that $\frac{\Delta N}{N} \sim \,- 10^{-3}$ to $-10^{-5}$.

\subsection{Effect of maximal acceleration in plasma wakefield acceleration and laser-electron acceleration}
The discrepancy in energy $\frac{\Delta E}{E_{\textrm{rel}}}$ depends upon the species of particle by the inverse of the maximal proper acceleration squared \eqref{effects depends upon the mass} and also on the characteristics of the accelerating system. Very powerful systems of acceleration are laser-plasma and wake plasma acceleration, where the acceleration $a$ can be much higher than in RF-cavity systems. Indeed, current laser electron acceleration systems achieve acceleration of bunches of electrons up to proper accelerations of order $10^{22}\, m/s^2$, which is much larger than the accelerations at LHC.

In order to illustrate this concepts, let us discuss three significant examples of laser-plasma acceleration. In all our examples, the particles accelerated are electrons. This implies that the maximal acceleration for an individual electron is $A_{\textrm{max}}\approx \,10^{32} \, m/s^2$, of order $10^{3}$ times smaller than for an individual proton. In contrast, the population of the bunches reached by laser-plasma is smaller than the ones discussed for the LHC.

In the first example that we discuss, the plasma wakefield acceleration achieved an acceleration gradient of $4.4\,GV/m$ for a bunch of particles of $74\,pC$ \cite{Litos et al. 2014}. These figures are  equivalent to an acceleration for an electron of order $7.9\times \,10^{20}m/s^2$ and a bunch population of $N\approx \,\,4.9\times \,10^7$. Since $M=1$, the ratio $\frac{\Delta E}{E_{\textrm{rel}}}$ given by the expression \eqref{effects depends upon the mass} is of order $\frac{\Delta E}{E_{\textrm{rel}}}\sim 7\times 10^{-8}$.

The second example that we consider is the laser acceleration experiment by Kurz et al. \cite{Kurz et al. 2021}. They reported an acceleration gradient of $100 \,GV/m$, while the measured integrated charged is $12 \, pC$, that we interpret here as the bunch's charge. These figures are equivalent to an acceleration of individual electrons of order $a\approx 1.8\times \,10^{22}\, m/s^2$ and $N\approx \,5\times\, 10^{7}$. The discrepancy with the relativistic theory is then given by the ratio $\frac{\Delta E}{E_{\textrm{rel}}}\sim 4\times 10^{-5}$.

The third example that we consider is the experiment with laser electron acceleration discussed in \cite{Wang et al. 2013}. In this case, the acceleration gradient is of order $a\sim 2\,GV/cm$. The bunch's population is identified with the charge at peak, which is approximately $63 \,pC$. These figures are equivalent to an acceleration $a\sim\,3\times \,10^{22}\,m/s^2$, while the population of the bunch is $N\sim\,3.9 \,\times 10^{8}$. Performing the analogous estimates, we obtain the ratio for the experiment of \cite{Wang et al. 2013} an increase of the energy of the order $\frac{\Delta E}{E_{\textrm{rel}}}\sim 7\times 10^{-3}$.

From the above examples, we observe that for laser acceleration or plasma acceleration, the population $N$ of the accelerated bunch can be of order $N\sim \,10^8$. This is lower than in LHC bunches, but the huge acceleration reached and the fact that in our model, the maximal acceleration for an individual electron is of order $10^{-3}$ less than for a proton, the discrepancy $\frac{\Delta E}{E_{\textrm{rel}}}$ leads to figures higher than for RF-acceleration and that potentially can be even higher. Also the detection methods in laser plasma acceleration, based on spectroscopic identification of individual particles, are good adapted to characterize either the excess of individual energy or the defect of population of the bunches. This is particularly relevant for observation, since contrary to the LHC, the effect of population loss in the bunch happens for $M=1$.
\section{Conclusion and discussion}\label{Conclusion and Discussion}
In this paper we have investigated consequences of the theory of spacetimes with maximal acceleration as developed in \cite{Ricardo2015} on the mass and related effects. We have seen that the value of the maximal proper acceleration is of the form $3/2\,m_0/q^2$. This value is, except by a constant of order $1$, the value found by Caldirola in his theory of the electron, as a direct result of the {\it quantum of time} hypothesis \cite{Caldirola}. We did not make use of the chronon hypothesis, but a detailed analysis about the relation between these two concepts could be of certain theoretical interest.

The discrepancy in the number population $\frac{\Delta E}{E}$ has an {\it additive} character, due to the appearance of the factor $M$ in the relation \eqref{effects depends upon the mass}. Thus a natural way to increase the quotient $\frac{\Delta N}{N}$ is either increase the acceleration $a^2$ or increase the number of independent accelerations $M$ or increase the population bunch $N$.
 In the case of plasma or laser acceleration, $M=1$. However, the effect can be relatively large compared with conventional accelerations because the maximal proper acceleration for an electron is of order $10^3$ smaller than for a proton accelerator. Therefore, as long as laser-plasma acceleration can reach more populated bunches with the current top scales of acceleration, the effect discussed in this paper will be increasingly relevant and provide potential observable effect of the equation of motion \eqref{equationofmotion} and the maximal proper acceleration \eqref{valueofthemaximalacceleration}.

The space charge and beam current effects can pose potential problems for the consistency of the model of the bunch used in this paper. The effects of space charge as usually considered in conventional particle accelerators are constrained by the condition that a transverse focusing field dominates the de-focusing effect of the plasma space charge and current effects. Let us adopt a simple model for the bunch as an axially symmetric cylinder. A necessary condition for this to happen is that the frequency of the transverse motion is larger than the plasma frequency of the beam  \cite{Davidson Qin},
\begin{align}
s_b:=\,(1-\beta^2)\,\omega^{2}_{pb}/\omega^2_{\beta\bot}<1,
\label{focusing over defocusing condition}
\end{align}
 where $\omega_{pb}$ is the plasma frequency associated with the bunch of particles, considered as a nonneutral plasma, $\omega_{\beta\bot}$ is the frequency associated to the transverse motion of the particle and $1/(1-\beta^2)^{1/2}$ is the relativistic gamma factor for the axial speed of the average  beam $V_b$. The factor $(1-\beta^2)$ acts as a focusing factor. Thus when adopting the assumption of the stability of the bunch in our theory, we are also adopting the condition $s_b<1$. Assuming a cylindrical model for the bunch, the dominant order of the focusing external electromagnetic force acting on one particle is proportional to $\omega^2_{\beta\bot}$, while the de-focusing force is proportional to $(1-\beta^2)/\omega^2_{pb}$ \cite{Davidson Qin}. Thus the condition $s_b<1$ can be read as that the focusing external forces dominate over the self-interacting, de-focusing forces. In our model, the acceleration of one particle is equivalent to the acceleration of the whole bunch of particles considered as a sole particle of mass  $m_0\,N$ and charge $q\, N$. Therefore, as long as $s_b<1$ holds good, the estimates for the acceleration and its effects discussed in this paper provide reasonable results if one is looking for the order of magnitude of the effect of maximal proper acceleration.

 Furthermore, the limit when $s_b<<1$ provides a condition for visibility of the effects discussed in this paper compared with self-interactions effects. Such a condition can be achieved in the ultra-relativistic limit, since $s_b\to 0$ when $\beta\to 1$. In more general situations, more accurate models for the bunch than the one discussed in this paper are necessary. Fluid models or kinetic models for plasmas in spacetimes of maximal proper acceleration could provide significant improvement in the understanding of the dynamics of particle beams in spacetimes with maximal proper acceleration and the effects of maximal proper acceleration on charge beam dynamics.

\appendix

\section{Jet bundles and generalized tensors}\label{Appendix A}
The notion of a generalized higher order field and the generalized higher order electrodynamics is introduced in this {\it appendix}. For more details of the theories discussed the reader can consult \cite{Ricardo2012,Ricardo 2017}. For the notions of differential geometry of jets, the reader can consult \cite{KolarMichorSlovak}.

Let $M$ be a smooth manifold and let us consider a local coordinate system on $M$ containing the point $x\in M$.
Given a smooth curve $\vartheta:I\to { M}$ such that $\vartheta (0)=x$, the collection of coordinates and its derivatives
$(\vartheta^\mu(0),\,\frac{d\vartheta^\mu}{d t}|_0,\,\frac{d^2\vartheta^\mu}{dt^2}|_0,..., \frac{d^k\vartheta^\mu}{d t^k}|_0)$ determines a $k$-jet at the point $x\in M$.
 The jet bundle $J^k_0(M)$ over $M$ is the disjoint union
\begin{align*}
J^k_0(M):=\bigsqcup_{x\in M}\,J^k_0(x).
\end{align*}
The canonical projection map is
\begin{align*}
^k\pi  :J^k_0(M)\to M,\quad
 (\vartheta(0),\,\frac{d\vartheta}{d t}\big|_0,\,\frac{d^2\vartheta}{d t^2}\big|_0,..., \frac{d^k \vartheta}{d t^k}\big|_0)\mapsto \vartheta (x(0)).
\end{align*}
Relevant for our constructions will be the canonical lift of $\vartheta:I\to M$, $j^k \vartheta:I\to J^k_0(M)$, $t\mapsto(\vartheta(t),\vartheta'(t),...,\vartheta^{k}(t))\in \,J^k_0(M)$.

The following definition provides the fundamental notion of generalized higher order jet tensor and forms,
\begin{definicion}
A generalized tensor $T$ of type $(p,q)$, $p,q\geq 0$, with values on $\mathcal{F}(J^k_0(M))$ is a smooth section of the bundle $T^{(p,q)}(M,\mathcal{F}(J^k_0(M)))$ of $\mathcal{F}(M)$-linear homomorphisms
$Hom(T^*M\times...^p...\times T^*M\times TM\times ...^q...\times TM,\,\mathcal{F}(J^k_0(M)))$.

A generalized $p$-form $\omega$ with values on $\mathcal{F}(J^k_0(M))$ is a smooth section of the bundle $\Lambda^p(M,\mathcal{F}(J^k_0(M)))$ of $\mathcal{F}(M)$-linear completely alternate homomorphisms $Alt(TM\times ...^p...\times TM,\,\mathcal{F}(J^k_0(M)))$.

The space of $0$-forms is $\Gamma\,\Lambda^0(M,\mathcal{F}(J^k_0(M)))=:\mathcal{F}(J^k_0(M))$.
\label{definiciontensoerFJMvaluados}
\end{definicion}

The generalized exterior algebra and the associated Cartan's  calculus is developed  in close analogy with  the
standard Cartan's calculus of smooth differential forms. In local coordinates $(J^k_0 U, \,x^\mu,\frac{d{x}^\mu}{dt}|_{t=0},\frac{d{x}^\mu}{d t^2}|_{t=0},...)$ over the open domain $J^kU\subset \,J^k_0(M)$, the exterior derivative of a $1$-form is
\begin{align*}
d_4\phi & =d_4\big(\phi_i(x,\,{x}',\,{x}'',..., x^{(k)}) d_4x^i\big)\\
& = d\big(\phi_i(x,\,{x}',\,{x}'',..., x^{(k)})\big)\wedge\, d_4x^i\\
& =\partial_j\,\phi_i(x,\,{x}',\,{x}'',..., x^{(k)})\,d_4x^j\,\wedge d_4x^i.
\end{align*}
This operator is nil-potent, $(d_4)^2=0$. One can extend this operation to general $p$-forms in an analogous way as in standard Cartan's calculus \cite{BennTucker,KolarMichorSlovak}.
\begin{definicion}
A generalized metric $g$ is a generalized symmetric tensor of type $(0,2)$ with the regularity properties of a metric in pseudo-Riemannian geometry.
\end{definicion}
A generalized metric  determines a star  operator $\star_g$ acting on the algebra
 \begin{align}
 \Lambda^p(M,\mathcal{F}(J^k_0(M))):=\,\sum^n_{p=0}\oplus \Lambda^p(M, \mathcal{F}(J^k_0 (M)))
 \end{align}
  in the following way. Let $\{e^\mu(j^kx),\,\mu=1,...,n\}$ be a local, orthonormal frame respect of $g$ at the point $j^kx \in \,J^kM$.
   The Levi-Civita symbol is denoted by $\epsilon_{\mu_1... \mu_n}$. Then the $\star_g$ operator of the algebra
   $\Lambda^p(M, \mathcal{F}(J^k_0 (M)))$ is the $\mathcal{F}(J^k_0(M))$-multilineal map determined
   by the image on the elements $e^{\mu_1}\wedge\cdot\cdot\cdot \wedge e^{\mu_p}$,
\begin{align*}
\star_g :\Gamma\Lambda^p (M,\mathcal{F}(J^k_0(M)))\to \Gamma\Lambda^{n-p} (M,\mathcal{F}(J^k_0(M)))
\end{align*}
\begin{align}
& (e^{\mu_1}(\,^kx)\wedge\cdot\cdot\cdot \wedge e^{\mu_p}(j^kx))\mapsto \epsilon_{\nu_1...\nu_n}\,g^{\mu_1 \nu_1}\,
\cdot\cdot\cdot g^{\mu_p \nu_p}e^{\nu_{p+1}}(j^kx)\wedge\cdot\cdot\cdot \wedge e^{\nu_n}(j^kx)
\label{hodgestaroperator}
\end{align}
and then it is extended to an arbitrary generalized form
\begin{align*}
\alpha=\alpha_{\mu_1....\mu_p}(j^kx)\,e^{\mu_1}(j^kx)\wedge\cdot\cdot\cdot
\wedge e^{\mu_p}(j^kx)\,\in\Gamma \Lambda^p(M, \mathcal{F}(J^k_0 (M)))
\end{align*}
 by the multilineal property,
\begin{align*}
\star\,\alpha(j^kx)=\,\alpha_{\mu_1....\mu_p}(j^kx)\epsilon_{\nu_1...\nu_n}\,g^{\mu_1 \nu_1}\,
\cdot\cdot\cdot g^{\mu_p \nu_p}e^{\nu_{p+1}}(j^kx)\wedge\cdot\cdot\cdot \wedge e^{\nu_n}(j^kx).
\end{align*}
By direct computation in a orthogonal frame, one can prove the following generalization of the Hodge star operator as in Riemannian geometry.
\subsection{Generalized higher order jet electromagnetism}
Let $M_4$ be a smooth $4$-manifold.
\begin{definicion}
The electromagnetic field $\bar{F}$ along the lift $j^k x:I\to J^k_0(M_4)$ is a $2$-form  that in local natural coordinates can be written as
\begin{equation}
\bar{F}(j^kx)=\bar{F}(x,{x}',{x}'',{x}''',..., x^{(k)})=\big(F_{\mu\nu}(x)+
\Upsilon_{\mu \nu}(x,{x}',{x}'',{x}''',..., x^{(k)})\big)d_4 x^{\mu}\wedge d_4 x^{\nu}.
\label{electromagneticfield}
\end{equation}

The excitation tensor $\bar{G}$ along the lift $j^kx:I\to J^k_0(M_4)$ is a $2$-form
\begin{equation}
\bar{G}(j^kx)=\bar{G}(x,{x}',{x}'',{x}''',..., x^{(k)})=\big(G_{\mu\nu}(x)+
\Xi_{\mu \nu}(x,{x}',{x}'',{x}''',..., x^{(k)})\big)d_4 x^{\mu}\wedge d_4 x^{\nu}.
\label{excitationtensor}
\end{equation}

The density current $\bar{J}$ is represented by a $3$-form
\begin{equation}
\bar{J}(x,{x}',{x}'',{x}''',..., x^{(k)})=\,\big(J_{\mu\nu\rho}(x)\,+
\Phi_{\mu\nu\rho}(x,{x}',{x}'',{x}''',..., x^{(k)})\big)\,d_4x^{\mu}\wedge d_4x^{\nu}\wedge d_4x^{\rho}.
\label{generalizedcurrent}
\end{equation}
\end{definicion}
 To each of these three types of generalized fields there are associated standard fields $F,G\in \,\Lambda^2 M_4$ and $J\in\,\Lambda^3M_4$ defined locally by the components $F_{\mu\nu}(x)$, $G_{\mu\nu}(x)$ and $J_{\mu\nu}(x)$ respectively.

The generalized homogeneous of Maxwell's equations are of the form
\begin{align}
d_4\bar{F}=0.
\label{homogeneousequation}
\end{align}
If we assume the constitutive relation $\bar{G}=\,\star \bar{F}$, then the generalized inhomogeneous Maxwell's equations are
\begin{align}
d_4\star \,\bar{F}=\,{J}+d_4\,\star \Upsilon.
\label{nonhomogeneous equation}
\end{align}

From equations \eqref{homogeneousequation} and \eqref{nonhomogeneous equation}
it is possible to construct an effective theory which is equivalent to the standard Maxwell's theory. In particular, the Faraday tensor $F$ must be such that
\begin{align}
d F=0
\label{equationforF}
\end{align}
holds good; $G$ must be a solution of the equation
\begin{align}
d\star\, F=\,J
\label{equationfor*F}
\end{align}
and the conservation of the current density are
\begin{align}
d{J}=0.
\label{equationforJ}
\end{align}

\section{Notion of Spacetime of maximal proper acceleration}\label{Appendix B}

Let $(M_4,\eta)$ be a Lorentzian structure.
  Let $\nabla:\Gamma TM_4 \times\to \Gamma TM_4$ be the covariant derivative associated to the Levi-Civita connection of $\eta$.
\begin{proposicion}
There is a non-degenerate, symmetric form $g$ along a curve $\vartheta:I\to M_4$  defined by the expressions:
\begin{itemize}
\item If $\eta(\vartheta',\vartheta')\neq 0$, then
\begin{align}
g(j^2\vartheta) (\vartheta',\vartheta')=\Big(1+ \frac{  \eta(\nabla_{\vartheta'}\vartheta'(t),
\nabla_{\vartheta'}\vartheta'(t))}{A^2 _{\textrm{max}}\,\eta(\vartheta',\vartheta')}\Big)\,\eta(\vartheta',\vartheta');
\label{maximalaccelerationmetric0}
\end{align}
\item If $\eta(\vartheta',\vartheta')=\,0$ holds, then $g(\vartheta',\vartheta')=\,\eta(\vartheta',\vartheta')$.
\end{itemize}
\label{teoremasobremaximaacceleration}
\end{proposicion}
The bilinear form $g$ determined by {\it Proposition} \ref{teoremasobremaximaacceleration} is the metric of maximal proper acceleration. Its action on two arbitrary vector fields $W,Q$ along $\vartheta:I\to M_4$ is given by the pointwise application of the expression \eqref{maximalaccelerationmetric}.
 Note that $g$ is not bilinear on the {\it base point} $\vartheta'(t)$ but it is bilinear on the vector arguments $W$, $Q$.

The metric of maximal proper acceleration is a generalized tensor of order $(0,2)$ of the form
 \begin{align}
 g(j^2\vartheta)=\,g^0(\,\vartheta,\vartheta',\vartheta'')+\,g^1(x,\vartheta',\vartheta'')\xi(x,\vartheta',\vartheta'',A^2_{\textrm{\textrm{max}}}).
  \label{perturabativeexpansion}
 \end{align}
We require that the limit
$\lim_{A^2_{\textrm{max}}\to +\infty}\,g(j^2\vartheta)$
 to be compatible with the clock hypothesis, that implies that $g^0(\vartheta,\vartheta',\vartheta'')=g^0(\vartheta,\vartheta').$
Moreover, the metric $g^0(\vartheta',\vartheta')$ is non-degenerate, since $g$ is non-degenerate. $g^0$ is also symmetric and bilinear.
The expression \eqref{maximalaccelerationmetric} is recovered with the identifications
\begin{align*}
g^0 =\,\eta,\quad  g^1 =\,\eta,\quad \xi=\,\frac{  \eta(\nabla_{\vartheta'}\vartheta'(t),
\nabla_{\vartheta'}\vartheta'(t))}{A^2 _{max}\,\eta(\vartheta',\vartheta')}.
\end{align*}

We can now define the notion of spacetime of maximal proper acceleration,
\begin{definicion}
A spacetime of maximal proper acceleration is a pair $(M_4,g)$ where $M_4$ is a four-dimensional manifold and $g$ is a generalized metric tensor specified by {\it proposition} \eqref{teoremasobremaximaacceleration}.
\end{definicion}

We remark that as long as the condition $g(\nabla_{\dot{\vartheta}}\dot{\vartheta},\nabla_{\dot{\vartheta}}\dot{\vartheta})<\,A^2_{\textrm{max}}$, the inverse $g^{-1}$ is well defined. This provides the possibility to construct the Christoffel symbols of $g$ and hence a linear Berwald type connection of $g$, as discussed in detail in \cite{Ricardo 2020}.
\section{The equation of motion for a point charged particle in higher order jet electrodynamics}\label{Appendix C}
In this {\it appendix}, a summarized version of the derivation of the equation of motion \eqref{equationofmotion} is discussed. Dot-notation refers to the derivative with respect to the proper time of the metric $\eta$, while prime notation means derivative with respect to the proper time of the metric of maximal acceleration $g$ given by \eqref{propertime g}.
\subsection{Preliminary kinematical considerations}
The following relations hold good,
 \begin{align*}
 \eta(\dot{x},\dot{x})& =\,(1-\epsilon)^{-1}\,g(\dot{x},\dot{x})=\,(1-\epsilon)^{-1}\,g((1-\epsilon)^{1/2} {x}',(1-\epsilon )^{1/2}{x}')=\,g({x}',{x}'),
 \end{align*}
 where $\epsilon=\frac{\eta(\ddot{x},\ddot{x})}{A^2_{\textrm{max}}}$. We also have the following kinematic relations,
 \begin{align}
& g(\dot{x},\dot{x})=-1,\label{covariantkineticconstrain1}\\
& g(\dot{x},\ddot{x})=\,-\frac{\dot{\epsilon}}{2}+\,\mathcal{O}(\epsilon^2) \label{covariantkineticconstrain2}.
\end{align}

Since the theory of higher order jet electrodynamics is formally analogous to Maxwell's theory, then by a similar procedure as in the standard theory, Larmor's covariant radiation law holds in the form
\begin{align}
\frac{d{p}^{\mu}_{rad}}{d\tau}=-\,\left( 1+c_1\,\epsilon^\alpha+\cdot\cdot\cdot\right) \frac{2}{3}\,q^2 (\ddot{x}^{\rho}\,\ddot{x}^{\sigma}g_{\rho\sigma})(\tau)\,\dot{x}^{\mu}(\tau) .
\label{generalizedLarmor}
\end{align}

\subsection{Derivation of the new equation of motion of a point charged particle}

We assume a form of generalized Lorentz equation, where instead of a Maxwell field we have a generalized Maxwell field as a source of a generalized Lorentz force,
\begin{align*}
m_b(\tau)\,\ddot{x}^{\mu} & =\,q\,\bar{F}^{\mu}_\nu\,\dot{x}^\nu=\,q\,F^{\mu}\,_{\nu}\,\dot{x}^\nu+ \big(B^{\mu}\dot{x}_{\nu}-\,\dot{x}^{\mu}B_{\nu}\big)\dot{x}^{\nu}\\
& +\big(C^{\mu}\ddot{x}_{\nu}-\,\ddot{x}^{\mu}C_{\nu}\big)\dot{x}^{\nu} +\big(D^{\mu}\dddot{x}_{\nu}-\,\dddot{x}^{\mu}D_{\nu}\big)\dot{x}^{\nu}\,+...,
\end{align*}
 with $F^{\mu}\,_{\nu}=\,g^{\mu\rho}F_{\rho\sigma}.$ $m_b(\tau)$ is the bare mass parameter, that depends on $\tau$-time parameter.
 On the right hand side of the above expression all the contractions that appear in expressions as
 $\big(B^{\mu}\dot{x}_{\nu}-\,\dot{x}^{\mu}B_{\nu}\big)\dot{x}^{\nu}$, etc...  are performed with the metric $g$.
By using an analogous method to Cartan's moving frame, the generalized Lorentz force equation is cast in the form
\begin{align*}
m_b(\tau)\,\ddot{x}^\mu =\, q\,F^\mu\,_\nu\,\dot{x}^\nu\,+\beta_2\,\ddot{x}^\mu\,\dot{x}^\nu\,\dot{x}_\nu-\,\dot{x}^\mu \,
\beta_2\,\ddot{x}_\nu\,\dot{x}^\nu.
\end{align*}
Using the kinetic relations for $g$, one obtains the expression
\begin{align*}
m_b(\tau)\,\ddot{x}^{\mu} =\,q\,F^{\mu}\,_{\nu}\dot{x}^\nu-\beta_2 \ddot{x}^{\mu}-\beta_2\,\left(
-\frac{1}{2}\dot{\epsilon}+\,\textrm{higher order terms}\right)\dot{x}^{\mu},
\label{equationbeforerenormalization}
\end{align*}
where we have not written higher orders in $\epsilon$ and its derivatives terms.

The relation \eqref{equationbeforerenormalization} must be consistent with the generalized covariant Larmor's power radiation formula \eqref{generalizedLarmor}. Therefore, we require
\begin{align}
&\beta_2=\, \frac{4}{3}q^2\,\ddot{x}^\nu\,\ddot{x}_\nu\,\left(
-\frac{1}{2}\dot{\epsilon}+\,\textrm{higher order terms}\right)^{-1}.
\end{align}

The next step is to consider the re-normalization of the bare mass $m_b$,
\begin{align}
m_0=\, m_b(\tau)+\,\frac{4}{3}\,q^2\,\ddot{x}^\nu\,\ddot{x}_\nu\,\left(
-\frac{1}{2}\dot{\epsilon}+\,\textrm{higher order terms}\right)^{-1},
\label{renormalizationofmass}
\end{align}
a well defined expression, except in a set of measure zero. The mass parameter $m_0$ does not depend on the time parameter and is identified with the inertial mass of the charged point particle. This procedure leads to the expression \eqref{equationofmotionnoncovariant expression}, which is a second order ordinary differential equation respect to the proper parameter of $\eta$ for the point charged particle. At leading order in $a^2/A_{\textrm{max}}$, the covariant form  of \eqref{equationofmotionnoncovariant expression} is the equation \eqref{equationofmotion}.

 Note that the above derivation is restricted to the regimen when $\epsilon\ll 1$. However, we assume an extrapolation of the validity domain of the equation to other domains of larger acceleration, just limited by $A_{\textrm{max}}$. This is reasonable, since the equation \eqref{equationofmotionnoncovariant expression} does not have pre-accelerated or run-away solutions \cite{Ricardo 2017}.

\footnotesize{
}

\end{document}